\documentclass[debug,figure]{epl}
\usepackage{times}
\usepackage{amsmath}
\usepackage{amssymb}
\usepackage{amsthm}
\usepackage{amsfonts}
\usepackage{verbatim}

\DeclareFontFamily{OT1}{eusm}{} \DeclareFontShape{OT1}{eusm}{m}{n}
{<5> <6> <7> <8> <9> <10> <11> <12> <14.4> eusm10}{}
\DeclareMathAlphabet{\eusm}{OT1}{eusm}{m}{n}

\newcommand{\CalA}{{\mathcal A}}
\newcommand{\BB}{{\mathcal B}}
\newcommand{\CC}{{\mathcal C}}
\newcommand{\OO}{{\mathcal O}}
\newcommand{\R}{{\mathbb R}}
\newcommand{\Z}{{\mathbb Z}}
\newcommand{\eusmH}{{\eusm H}}
\newcommand{\eusmP}{{\eusm P}}

\newcommand{\Deltac}{\Delta_{\text{\rm c}}}
\newcommand{\lambdac}{\lambda_{\text{\rm c}}}
\newcommand{\betac}{\beta_{\text{\rm c}}}
\newcommand{\tauW}{\tau_{\text{\rm W}}}

\newcommand{\diam}{\mbox{\rm diam}\,}

\title{On the formation/dissolution of equilibrium droplets}
\shorttitle{Equilibrium droplets}
\author{M.~Biskup\inst{1} \and L.~Chayes\inst{1} \and R.~Koteck\'y\inst{2}}
\shortauthor{M.~Biskup \etal}
\institute{
  \inst{1} Department of Mathematics, UCLA, Los Angeles, California 90095-1555, USA\\
  \inst{2} Center for Theoretical Study, Charles University,
Jilsk\'a~1, 110~00 Prague, Czech Republic
}

\begin{document}

\maketitle

\begin{abstract}
We consider liquid-vapor systems in finite volume $V\subset\R^d$
at parameter values corresponding to phase coexistence and study droplet
formation due to a fixed excess $\delta N$ of particles above the
ambient gas density. We identify a dimensionless parameter
$\Delta\!\sim\!(\delta N)^{(d+1)/d}/V$ and a \textrm{universal}
value $\Deltac=\Deltac(d)$, and show that a droplet of the
dense phase occurs whenever $\Delta\!>\!\Deltac$, while,
for~$\Delta\!<\!\Deltac$, the excess is entirely absorbed into the
gaseous background. When the droplet first forms, it comprises a
non-trivial, \textrm{universal} fraction of excess particles. 
Similar reasoning applies to generic two-phase systems at phase coexistence including solid/gas---where the ``droplet'' is crystalline---and polymorphic~systems. A
sketch of a rigorous proof for the 2D~Ising lattice gas is
presented; generalizations are discussed heuristically.
\end{abstract}

\section{Introduction}
The thermodynamics of droplets in systems with phase
coexistence has been well understood since the pioneering
works~\cite{Gibbs,Curie,W,Langmuir}. Recently, justifications of
the classic results based on the first principles of statistical
mechanics have been attempted---in both
two~\cite{ACC,DKS,DS,Bob+Tim,PV} and higher~\cite{Cerf,Bodineau,Cerf-Pisztora,BIV}
dimensions---and various thermodynamical predictions concerning
macroscopic shapes have been rigorously established. However, the
\textit{formation} and \textit{dissolution} of equilibrium
droplets is among the less well-studied areas in statistical
mechanics. Indeed, most of the aforementioned analysis has focused
on the situation implicitly assumed in the classical derivations;
namely, that the scale of the droplet is comparable with the scale
of the system. As is known~\cite{DS,Bob+Tim,Binder-Kalos,Sethna,nemci}, this will not
be the case when the parameter values are such that a droplet
first forms. In this Letter, we underscore the region of the
system parameters that is critical for the formation/dissolution
of droplets. In particular, we isolate the mechanism by which the
low-density phase copes with an excess of particles and pinpoint
the critical amount of extra particles needed to cause a droplet
to appear. Surprisingly, at the point of droplet formation, only a
certain fraction of the excess goes into the droplet; the rest is
absorbed by the bulk. Moreover, apart from a natural rescaling to
dimensionless parameters, all of the above can be described in
terms of universal quantities independent of the system
particulars and the temperature.

In the last few years, there has been some interest in questions
related to droplet formation and dissolution with purported
applicability in diverse areas such as nuclear
fragmentation~\cite{fragment1,fragment2,fragment3} and the stability of adatom islands
on crystal surfaces~\cite{MS,Sethna}. Another issue, which is of
practical significance in statistical mechanics, concerns the
detection of first-order phase transitions by the study of small
systems with fixed order parameter (magnetization) or fixed
energy. Under these conditions, non-convexities appear in the
finite-volume thermodynamic functions (which, of course, must
vanish in the thermodynamic limit),
see~\cite{G,PH,Chinese,Kosterlitz}. Naturally, this suggests the
formation of a droplet in a system with coexisting phases. Several
studies have directly addressed the issues surrounding the
appearance of droplets with intriguing reports on finite-size
characteristics~\cite{Farukawa-Binder,PS,PH,MS,Sethna}. We believe that the
results of this Letter may shed some light in these situations.

\section{Droplets in systems at phase coexistence}
We will start with some general considerations which bolster the
claims of the first paragraph and, at the end of this Letter,
describe the principal steps of a rigorous proof for the~2D Ising
lattice gas. Although the natural setting for these problems is
the canonical distribution in finite volume, intuition is often
better developed in the context of finite subsystems using the
language of the grandcanonical ensemble. Here the occurrence of
droplets may be regarded as a problem in large deviation theory.
This perspective will guide our heuristic analysis as it did in
the proof for the 2D~Ising~system.

Consider a generic liquid-vapor system.
(In this Letter, we adopt, for concreteness, the language of the
liquid-vapor transition. However, all considerations apply equally
well to the formation/dissolution of an equlibrium crystal against a liquid or
gaseous background.)
First, suppose that the
system is in the gaseous phase. According to a
fluctuation-dissipation analysis, the local fluctuations for
subsystems of volume $V$ are then of the order $\sqrt{\varkappa
V}$, where, modulo constants, $\varkappa$ is the isothermal
compressibility. More precisely, the probability of observing a
particle excess $\delta N$ is given by
\begin{equation}
\label{Gauss} \exp\Bigl\{-\frac{\,(\delta N)^2}{2\varkappa
V}\Bigr\}.
\end{equation}
Now, when $\delta N=\OO(\sqrt{\varkappa V})$, the above is just
the leading-order asymptotic of a full-fledged Gaussian (central
limit) distribution, which comes equipped with power-law
corrections, etc. Moreover---in the single phase regime---the
above leading order remains valid even for $(\delta N)^2\gg
\varkappa V$, provided that $|\delta N|\ll \rho_{\text{G}}V$,
where $\rho_{\text{G}}$ is the gas density.

In the two-phase regime, small excesses can be again absorbed into
background fluctuations but, in addition, a second mechanism
exists through which the system can handle an excess of particles;
namely, the formation of liquid-phase droplets. The minimal cost
of a droplet of volume $\delta V$ goes as
\begin{equation}
\label{Wulff} \exp\bigl\{-\tauW(\delta V)^{\frac{d-1}d}\bigr\},
\end{equation}
where $\tauW$ denotes the (surface) free energy of an ideal-shape
droplet of unit volume. For an isotropic system, $\tauW=\tau S_d$,
where $\tau$ is the surface tension and
$S_d=2\pi^{d/2}/\Gamma(\frac d2)$ is the surface area of the unit
sphere in~$\R^d$. In general, $\tauW$ is obtained by minimizing
the Wulff functional~\cite{W}. As noted already in \cite{Gibbs,Curie,W},
by isoperimetric inequalities, scenarios 
involving 
multiple macroscopic droplets are far less likely.

Now the number of \textit{excess} particles $\delta N$ in a
droplet of~volume $\delta V$ is just $\delta
N=(\rho_{\text{L}}-\rho_{\text{G}})\delta V$, where
$\rho_{\text{G}}$ and $\rho_{\text{L}}$ are the ambient gas and
liquid densities, respectively. Thus,~comparing Eqs.~(\ref{Gauss})
and~(\ref{Wulff}), the droplet mechanism dominates when $\delta
N\gg \Theta V^{d/(d+1)}$, where
$\Theta^{d+1}=(\varkappa\tauW)^d(\rho_{\text{L}}-\rho_{\text{G}})^{1-d}$,
while the fluctuation mechanism dominates when $\delta N\ll\Theta
V^{d/(d+1)}$. We note that, from the perspective of rigorous
analysis, significant progress has been made in the single-phase
regime and in the above-mentioned extreme cases of the two-phase
regime. For the Ising model, this was done for low temperatures in
the exhaustive paper~\cite{DS}, while~\cite{Bob+Tim} extended this
result throughout the coexistence region in the case $d=2$.

Previously, the crossover region $\delta N\approx\Theta
V^{d/(d+1)}$ has not received adequate attention. To study this
region, we introduce the appropriate dimensionless parameter
\begin{equation}
\label{Delta}
\Delta=\frac{\,\,(\rho_{\text{L}}-\rho_{\text{G}})^{\frac{d-1}d}}
{2\varkappa\tauW}\frac{\,\,(\delta N)^{\frac{d+1}d}}V
\end{equation}
and investigate finite (but large) size systems as $\Delta$
varies. The key to the whole picture is that, for the entire range
of $\Delta$, there is a forbidden interval of droplet sizes. To be
precise, let us categorize droplets according to their surface
area: We will say a droplet is of \textit{intermediate} size if
its surface area is large compared with $\log V$ but small
compared with $V^{(d-1)/(d+1)}$. Droplets with surface areas
outside this range will be called \textit{large} and
\textit{small} as appropriate. We will show that, with
overwhelming probability, there are no intermediate droplets.

We begin with some observations: Suppose we specify the amount of
excess which goes into intermediate and large scale droplets and
fix the location of these droplets. Then, throughout the rest of
the system, the fluctuations-dissipation result in
Eq.~(\ref{Gauss}) is valid, at least to leading order. Indeed, the
only obstructions to Gaussian-type bulk fluctuations are: (1)~The
appearance of droplets beyond the logarithmic scale, which we have
already separated for a special treatment, and (2)~An exorbitant
surface to volume ratio. Since $\delta N\approx\Theta V^{d/(d+1)}$
(i.e., $\Delta<\infty$), the second possibility does not occur.~Let
\begin{equation}
\label{eq4} 
\delta N=\delta N_{\text{L}}+\delta
N_{\text{I}}+\delta N_{\text{S}},
\end{equation}
where $\delta N_{\text{L}}$ is the amount of excess particles in
large droplets, and similarly for $\delta N_{\text{I}}$ and
$\delta N_{\text{S}}$. By Eq.~(\ref{Gauss}), the distribution of
the excess given in Eq.~(\ref{eq4}) has a cost $\exp\{-(\delta
N_{\text{S}})^2/(2\varkappa V)\}$ for the fluctuation part. The
cost of intermediate droplets will be of the order of their
combined surface. If there are~$n$ such droplets, then
isoperimetric reasoning forces us to pay at least
$\exp\{-\tauW C(\delta N_{\text{I}})^{(d-1)/d}n^{1/d}\}$, where
$C$ is a constant of order unity. On the other hand, if all of
$\delta N_{\text{I}}$ were to go into small scale fluctuations, we
would simply have to pay $\exp\{-(\delta N_{\text{S}}+\delta
N_{\text{I}})^2/(2\varkappa V)\}$. Comparing the two mechanisms we
find, using $\delta N_{\text{S}}\lesssim \Theta V^{d/(d+1)}$ and
$\delta N_{\text{I}}\ll n \Theta V^{d/(d+1)}$, that
\begin{equation}
\label{bound} \frac{(\delta N_{\text{S}})^2}{2\varkappa V}+\tauW
C(\delta N_{\text{I}})^{\frac{d-1}d}n^{1/d}\gg \frac{(\delta
N_{\text{S}}+\delta N_{\text{I}})^2}{2\varkappa V},
\end{equation}
whenever $n\ge1$.
Hence, the probability of even a single droplet
of the intermediate scale is utterly negligible.

Having established the absence of intermediate-scale droplets,
isoperimetric inequalities rule out the possibility of more than
one large droplet. Thus, we are down to the simplest possible
scenario: There is (at most) a single large droplet in the
system---the cost of which is governed by
Eq.~(\ref{Wulff})---absorbing some of the excess, while the rest
goes into background fluctuations---which are described by
Eq.~(\ref{Gauss}). Thus, the probability that the droplet contains
the fraction $\lambda$ of the excess particles is, in the leading
order, given by
\begin{equation}
\exp\biggl\{-\tauW\,\Bigl(\frac {\delta N}
{\rho_{\text{L}}-\rho_{\text{G}}}\Bigr)^{\!\!\textstyle\frac{d-1}d}
\,\Phi_\Delta(\lambda)\biggr\},
\end{equation}
where $\Phi_\Delta(\lambda)$ is defined by
\begin{equation}
\Phi_\Delta(\lambda)=\lambda^{\frac{d-1}d}+\Delta (1-\lambda)^2.
\end{equation}
In particular, with overwhelming probability, the fraction of
excess particles taken by the droplet corresponds to a value
of~$\lambda$ that minimizes $\Phi_\Delta(\lambda)$.

\begin{figure}[t]
\onefigure[width=3.5in]{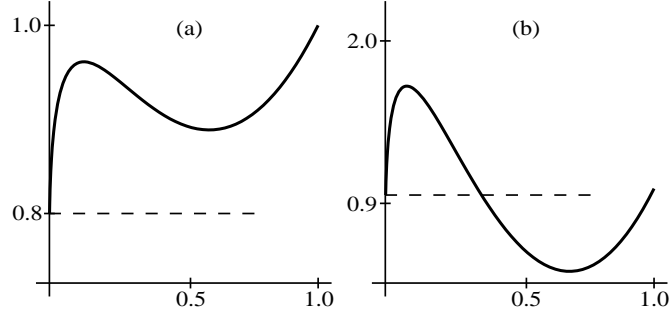}
\caption[Fig1]{The graph of the universal function
$\Phi_\Delta$ in $d=2$. Here the parameter $\lambda$ represents 
the trial fraction of the excess that goes into the droplet; $\Phi_\Delta$ has the interpretation of a \textit{free  energy} function. In
(a),~$\Delta=0.8<\Deltac$, and the function is minimized by $\lambda=0$. In (b),~$\Delta=0.96>\Deltac$ and the function is minimized by a $\lambda=\lambda_\Delta>2/3$. The maximum that interdicts between $\lambda=0$ and $\lambda=\lambda_\Delta$ presumably plays the role of a free energy barrier for
the formation of the droplet.}
\label{Fig1}
\end{figure}

\begin{figure}[t]
\onefigure[width=2.6in]{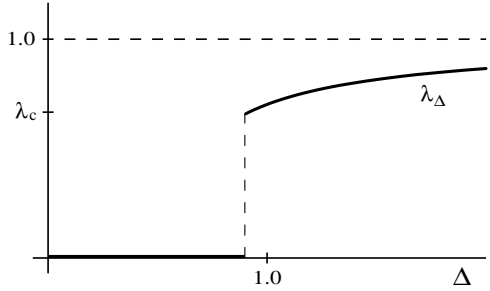}
\caption[Fig2]{The graph of $\lambda_\Delta$, the fraction of the excess that goes into the droplet, as a function of $\Delta$, the dimensionless rescaled parameter that measures the total excess. Here $d=2$. Notice that $\lambda_\Delta=0$ for $\Delta<\Deltac\approx0.918$, but as $\Delta\downarrow\Deltac$, $\lambda_\Delta$ tends to $\lambdac=2/3$. The behavior of the system 
at $\Delta=\Deltac$ has not been fully elucidated.}
\label{Fig2}
\end{figure}

The result of a straightforward computation is that there is a
constant $\Deltac$, given by the expression
\begin{equation}
\label{Delta-c}
\Deltac=\frac1d\Bigl(\frac{d+1}2\Bigr)^{\!\!\textstyle\frac{d+1}d},
\end{equation}
which separates two types of behavior: For~$\Delta<\Deltac$, the
unique global minimizer of $\Phi_\Delta(\lambda)$ is $\lambda=0$,
while for $\Delta>\Deltac$, the unique global minimum of
$\Phi_\Delta$ occurs at a non-trivial value $\lambda_\Delta>0$,
see Fig.~\ref{Fig1}. Moreover, the quantity $\lambda_\Delta$
increases monotonically with $\Delta$ and the value of
$\lambda_\Delta$ at $\Delta=\Deltac$, denoted by $\lambdac$, can
be computed exactly;
\begin{equation}
\lambdac=\frac2{d+1}.
\end{equation}
In particular, we have $\lambda_\Delta\ge\lambdac$ for all
$\Delta\ge\Deltac$. See Fig.~\ref{Fig2}.

Let us interpret the results in the context of the canonical
distribution: The region $\Delta<\Deltac$ minimized by
$\lambda\equiv0$ is the remnant of the ``phase'' $\delta N\ll
\Theta V^{d/(d+1)}$; the entire excess is taken up by background
fluctuations. For $\Delta>\Deltac$, a large droplet occurs which
absorbs the fraction $\lambda_\Delta$ of the particle excess.
Although this is obviously a precursor to the droplet-dominated
``phase'' (where $\delta N\gg\Theta V^{d/(d+1)}$), the physics is
somewhat different since a finite fraction of the excess---namely
$(1-\lambda_\Delta)\delta N$ particles---is still handled by the
background. We emphasize that $\lambda_\Delta$ and $\Delta$ are
related via a simple algebraic equation. The system-specific
details and dependence on external parameters are encoded into the
factor $(\rho_{\text{L}}-\rho_{\text{G}})^{(1-d)/d}\varkappa\tauW$
from Eq.~(\ref{Delta}); the dimensionless
parameter~$\lambda_\Delta$ is a \textit{universal} function of
$\Delta$.

We remark that in~\cite{Binder-Kalos,Sethna,nemci}, 
similar conclusions had been reached by various circuitous routes 
under the mantel of specialized assumptions or approximations. In this note, the
exact formula has been derived on the basis of
simple-minded droplet/fluctuation-dissipation arguments, all of
which can be rigorously proved in at least one case.

\section{Mathematical results for 2D~Ising model}  In the context of the
two dimensional Ising lattice gas, the above reasoning has been
elevated to the status of a mathematical theorem, which for convenience we state
in the language of the equivalent spin system. Consider the
square-lattice Ising model with the (formal) Hamiltonian
\begin{equation}
\eusmH=- \sum_{\langle x,y\rangle} \sigma_x\sigma_y,
\end{equation}
where $\sigma_x=\pm1$ and $\langle x,y\rangle$ denotes a
nearest-neighbor pair. For each inverse temperature $\beta$, let
$m^\star=m^\star(\beta)$ be the spontaneous magnetization,
$\chi=\chi(\beta)$ the magnetic susceptibility, and
$\tauW=\tauW(\beta)$ be the minimal value of the Wulff functional
for droplets of unit volume. As is well known, $m^\star(\beta)>0$,
$0<\chi(\beta)<\infty$ and $\tauW(\beta)>0$ once
$\beta>\betac=\frac12\log(1+\sqrt 2)$.

Consider now an $L\times L$ square in $\Z^2$ denoted by
$\Lambda_L$ and let~$M_L=\sum_{x\in\Lambda_L}\sigma_x$ be the
overall magnetization in $\Lambda_L$. Let $v_L\ge0$ be such that
$m^\star|\Lambda_L|-2m^\star v_L$ is an allowed value of~$M_L$ for
all~$L$. Let $P_{L,v_L}^{+,\beta}$ be the canonical distribution
on~$\Lambda_L$ with plus boundary conditions, inverse
temperature~$\beta$, and~$M_L$ fixed to the value
$m^\star|\Lambda_L|-2m^\star v_L$. In the present setting, the
parameter $\Delta$ in~Eq.~(\ref{Delta}) becomes
\begin{equation}
\label{Delta-lim} \Delta=2\frac{(m^\star)^2}{\chi\tauW}
\,\lim_{L\to\infty}\frac{\,v_L^{3/2}}{|\Lambda_L|},
\end{equation}
where we presume that the limit exists.

\begin{figure}[t]
\onefigure[width=2.3in]{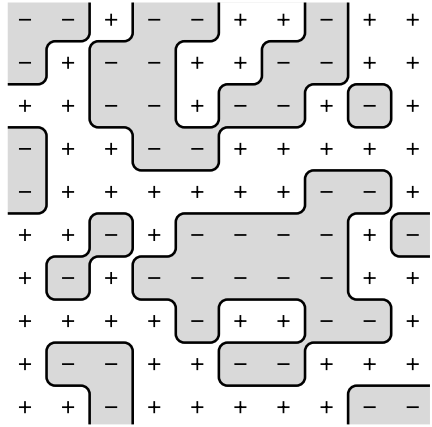}
\caption{Peierls' contours in a 2D-Ising
spin configuration. In the lattice-gas language, we regard 
minus spins as particles and plus spins as vacancies. 
Peierls' contours are then the interfaces separating 
high and low density regions at the microscopic scale.}
\label{Fig3}
\end{figure}

In Ising systems, a convenient description for the spin
configurations is in terms of their Peierls' contours, i.e., the
lines separating spins of opposite type; see Fig.~\ref{Fig3}.
Moreover, in the present context, the boundaries of droplets are
exactly these contour lines. Our first claim concerns the absence
of contours of intermediate size, regardless of the value
of~$\Delta$.

\smallskip\noindent
\textbf{Theorem I.}\ \ \textit{Let $\beta>\betac$ and suppose that the
limit in Eq.~(\ref{Delta-lim}) exists with $\Delta\in(0,\infty)$.
Let $\CalA_L$ be the event that there is \textit{no}
contour~$\Gamma$ in~$\Lambda_L$ with}
\begin{equation}
\label{forbidden} K\log L\le\diam\Gamma\le\frac1K L^{2/3}.
\end{equation}
\textit{If $K=K(\beta)$ is sufficiently large, then}
\begin{equation}
\lim_{L\to\infty}P_{L,v_L}^{+,\beta}(\CalA_L)=1.
\end{equation}

\smallskip
We remark that this is the rigorous (albeit 2D-Ising specific)
analogue of our general argument in
Eq.~(\ref{Delta})--Eq.~(\ref{bound}). The above theorem is far and
away the most difficult part of the mathematical analysis.
Notwithstanding, the flow of the proof parallels closely the
derivation that was given here. The reader may have noticed that,
in the present derivation, we have not used in any obvious way the
lower bound defining the scale of intermediate droplets. The issue
is somewhat delicate since some log-scale droplets will naturally
emerge from the background fluctuations. The constant $K$ must be
chosen large in order to enforce the distinction between
``natural'' and ``unnatural'' as well as to control the
translation entropy of the purported intermediate droplets.

Our next goal is to specify the typical configurations in measure
$P_{L,v_L}^{+,\beta}$ depending on the value~$\Delta$ as compared
with~$\Deltac$. Let $\BB_{L,K}$ be the event that there is
\textit{no contour $\Gamma$ in~$\Lambda_L$} with $\diam\Gamma\ge
K\log L$. Furthermore, let $\CC_{L,K}$ be the event that there is
\textit{one contour} $\Gamma_0$ with $\diam\Gamma_0\ge\frac1K
L^{2/3}$ while \textit{all other contours $\Gamma$ in $\Lambda_L$}
satisfy $\diam\Gamma\le K\log L$.

\smallskip\noindent
\textbf{Theorem II.}\ \textit{Let $\beta>\betac$ and suppose that the
limit in Eq.~(\ref{Delta-lim}) exists with $\Delta\in(0,\infty)$.
Let $\Deltac$ be as in Eq.~(\ref{Delta-c}) and, for
$\Delta>\Deltac$, let $\lambda_\Delta$ be the unique minimizer of
$\Phi_\Delta(\lambda)$.}

(1) \textit{If $\Delta<\Deltac$ and $K=K(\beta)$ is sufficiently large,
then}
\begin{equation}
\lim_{L\to\infty}P_{L,v_L}^{+,\beta}(\BB_{L,K})=1.
\end{equation}

(2) \textit{If $\Delta>\Deltac$ and $K=K(\beta)$ is sufficiently large,
then}
\begin{equation}
\lim_{L\to\infty}P_{L,v_L}^{+,\beta}(\CC_{L,K})=1.
\end{equation}
\textit{Moreover, with probability approaching one, the unique ``large''
contour $\Gamma_0$ has volume $(\lambda_\Delta+o(1))v_L$ and its
shape asymptotically optimizes the surface-energy (Wulff)
functional for the given volume.}

\smallskip
Theorems~I and~II completely classify the behavior of the Ising
system for all $\Delta\ne\Deltac$. We emphasize that the situation
at $\Delta=\Deltac$ has not been fully clarified. What can be
ruled out, according to Theorem~I, is the possibility of a
complicated scenario involving intermediate size droplets on a
multitude of scales. Indeed, when $\Delta=\Deltac$, in (almost)
every configuration we must have either a single large droplet or
no droplet at all; i.e., the outcome must mimic the case
$\Delta>\Deltac$ or $\Delta<\Deltac$. It is conceivable that one
outcome dominates all configurations or that both outcomes are
possible depending on auxiliary conditions.

Our last statement concerns the decay of the probability (in the
grandcanonical distribution) that the overall magnetization takes
value $m^\star|\Lambda_L|-2m^\star v_L$. Let $\eusmP_L^{+,\beta}$
be the (Gibbs) probability distribution on spins in $\Lambda_L$
with plus boundary condition and inverse temperature~$\beta$.

\smallskip\noindent
\textbf{Theorem III.}\ \ \textit{Let $\beta>\betac$ and suppose that
the limit in Eq.~(\ref{Delta-lim}) exists with
$\Delta\in(0,\infty)$. Introduce the shorthand
$p_L=\eusmP_L^{+,\beta}(M_L=m^\star|\Lambda_L|-2m^\star v_L)$.
Then}
\begin{equation}
\lim_{L\to\infty} v_L^{-1/2} \log p_L=-\tauW\inf_{0\le
\lambda\le1}\Phi_\Delta(\lambda).
\end{equation}

\smallskip
\section{Sketch of the proofs} 
Here we outline the steps
necessary to prove the above theorems. As already noted, first we
reduce the problem to the study of large-deviation properties of
the ``grandcanonical'' distribution $\eusmP_L^{+,\beta}$ using the
relation
\begin{equation}
P_{L,v_L}^{+,\beta}(\CalA)=\eusmP_L^{+,\beta}(\CalA|M_L
=m^\star|\Lambda_L|-2m^\star v_L),
\end{equation}
valid for all events $\CalA$. The next technical step is then a
proof of the large-deviation lower bound
\begin{equation}
\label{}
p_L\ge\exp\{-\tauW\sqrt{v_L}(\Phi_\Delta(\lambda)+\epsilon_L)\},
\end{equation}
which is produced by forcing in a contour of the appropriate size
and evaluating the contributions from surface tension and bulk
fluctuations. Here $\epsilon_L\to0$ as $L\to\infty$, uniformly
in~$\lambda$. A comparison with this lower bound then shows that,
with overwhelming probability in $P_{L,v_L}^{+,\beta}$, the total
surface area of contours $\Gamma$ with $\diam\Gamma\ge K\log L$ is
at most of order $\sqrt{v_L}$ while their combined volume is at
most of order~$v_L$. This puts us in a position to carry out the
argument in Eq.~(\ref{bound}), which ultimately leads to the proof
of Theorem~I. Having eliminated the intermediate contours, we are
down to the scenario with at most one large contour. Optimizing
over the contour volume/shape proves Theorem~II and also produces
a large-deviation upper bound, which completes the proof of
Theorem~III.

\smallskip
Theorems~I-III pretty much tell the story for this particular
case. Complete proofs and additional details will appear
elsewhere~\cite{BCK}.

\acknowledgments
The research of R.K. was partly supported by
the grants GA\v{C}R~201/00/1149 and MSM~110000001.
The research of L.C.~was supported by the NSF under the grant DMS-9971016 and
by the NSA under the grant NSA-MDA~904-00-1-0050.

\end{document}